\RequirePackage{fix-cm}
\documentclass[smallextended]{svjour3}       % onecolumn (second format)
\smartqed  % flush right qed marks, e.g. at end of proof
\usepackage{graphicx}

\usepackage{epstopdf}
\usepackage{natbib}
\usepackage{float}
\usepackage{url}
\usepackage{booktabs}
\usepackage{amsfonts}

 \journalname{Energy Systems}
\begin{document}

\title{A novel incentive-based demand response model for Cournot competition
	in electricity markets%\thanks{Grants or other notes
}

\author{Jos\'e Vuelvas         \and
        Fredy Ruiz %etc.
}

\titlerunning{Demand response and Cournot competition}        % if too long for running head

\institute{J. Vuelvas \at
	Departamento de Electr\'onica, Pontificia Universidad Javeriana,\\
	Bogot\'a, Colombia.\\
	\email{vuelvasj@javeriana.edu.co}           %  \\
	\and
	F. Ruiz \at
	Departamento de Electr\'onica, Pontificia Universidad Javeriana,\\
	Bogot\'a, Colombia.\\
	\email{ruizf@javeriana.edu.co}
}

\date{Received: date / Accepted: date}

\maketitle

\begin{abstract}
	This paper presents an analysis of competition between generators
	when incentive-based demand response is employed in an electricity
	market. Thermal and hydropower generation are considered in the
	model. A smooth inverse demand function is designed using a sigmoid
	and two linear functions for modeling the consumer preferences under
	incentive-based demand response program. Generators compete to sell energy bilaterally to
	consumers and system operator provides transmission and arbitrage
	services. The profit of each agent is posed as an optimization problem,
	then the competition result is found by solving simultaneously Karush\textendash Kuhn\textendash Tucker
	conditions for all generators. A Nash-Cournot equilibrium is
	found when the system operates normally and at peak demand times
	when DR is required. Under this model, results show that DR diminishes the energy consumption at peak periods, shifts the power requirement to off-peak times and improves the net consumer surplus due to incentives received for
	participating in DR program. However, the generators decrease their
	profit due to the reduction of traded energy and market prices.
\end{abstract}

\section{Introduction}

Demand Response (DR) is a program to motivate changes in electricity usage by customers
in response to changes in the price signal. DR is implemented
by system operator (SO) to match the load with power generation in a smart grid. Advanced metering at distribution side is required to implement a DR program \citep{Aketi2014}. The
main application is to decrease the load at peak times in
order to guarantee power availability and security on electrical grid \citep{Zhu2013,Bloustein2005,Su2009}. The aim is to control noncritical loads at residential, commercial and industrial levels for balancing
supply and demand. Broadly speaking, there are two ways to active DR: direct control (e.g. load shedding or set-point based solutions) \citep{Diaz2017} and indirect methods (such as price-based programs) \citep{Vuelvas2017}.  

There are some DR programs implemented as part of strategies to reduce
peak power. In \citep{Vardakas2015,Deng2015,Siano2014,Albadi2008,Madaeni2013},
some complete summaries regarding mathematical models, pricing
methods, optimization formulation and future extensions are described. 
An interesting program to induce DR is via incentive payment (an indirect control) by using
a technique called Peak Time Rebate (PTR) \citep{Vuelvas2017,Vuelvas2015,Mohajeryami2016,SeverinBorenstein2014}, where
customers receive electricity bill rebates by not consuming (relative
to a previously established, household-specific baseline) during peak
periods. In PTR, the baseline is a vital concept since the payment depends on the calculation of estimated consumption,
namely, a counterfactual model must be developed. In \citep{Mohajeryami2016},
some methods are explained to estimate the customer baseline. A randomized controlled trial method is developed  in order to establish customer baseline load, applied to aggregated forms of the consumption load in \citep{Mohajeryami2016b}. The critical facts on the selection of  customer baseline is shown in \citep{Chao2011}, authors design
a suitable baseline focusing on administrative and contractual approaches
in order to get an efficient DR. Furthermore, in \citep{Faria2013,Wijaya2014,Antunes2013},
the performance of DR baseline estimation models is studied and new methods are regarded
as establishing the reasonable compensation for the consumer. 

The co-existence of a variety of generation technologies
is an interesting problem from a gaming point of view and even more with the
integration of DR into the electricity market. In \citep{Genc2008}, the competition between
hydro and thermal electricity generators under uncertainty over demand
and water flows is presented. The authors in \citep{Garcia2005} analyze the price-formation
in an oligopoly model where hydroelectric generators are involved
in dynamic Bertrand competition. Furthermore, in \citep{Villar2003},
a model to understand a hydrothermal electric power market is built 
based on simple bids to the SO. Moreover, in \citep{Zhu2013},
by means of Stackelberg game is illustrated what is the value that
DR management can bring to generation companies
and consumers in a smart grid. In \citep{Su2009}, a method is devised 
for quantifying the effect of the demand response for the
market as a whole.  

\begin{figure}[h]
	\begin{centering}
		\includegraphics{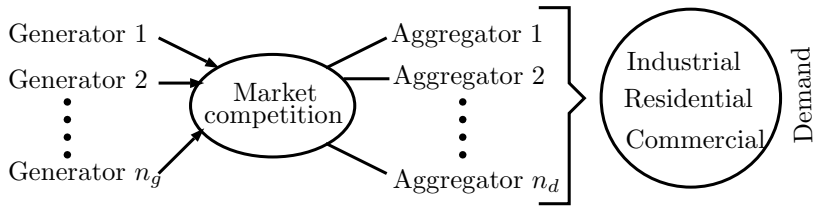}
		\par\end{centering}
	
	\caption{\label{fig:Electricity-market-with}Gaming in the electricity market. Where $n_g$ and $n_d$ are the total number of generators and aggregators in the electricity market, respectively.}
\end{figure}

The agents involved in an electricity market in competition with DR
are shown in Fig. \ref{fig:Electricity-market-with}. In this
paper, SO is responsible for arbitrage services in order to establish
a proper environment for competition and gaming. The generators have
different technologies, costs, revenues, and each firm seeks
to maximize its profit (the difference
between producers\textquoteright{} revenue and costs). Furthermore,
the aggregators carry out the request to users of reducing energy
consumption, namely, DR process. The main goal is to estimate the equilibrium
price under gaming environment. This competition is less
than perfect, some firms are able to influence the market price through
their actions. Such optimization problems set up which is
called in game theory a non-cooperative game \citep{VegaRedondo2003,Gabriel2013,Tirole1988,Osborne1995,Varian1992}.
The solution of such a game is called a Nash equilibrium and represents
a market equilibrium under imperfect competition. 

In this paper, a game among generators with different technologies in an electricity market is analyzed if
DR is required when the demand side exceed a defined threshold a priori by SO in order to guarantee some objectives of a smart grid. This threshold is determined from all customer baseline load and desired energy reduction during peak times. Then, a novel demand curve is proposed in order to understand the effect of electricity market behavior when an incentive-based DR program, like PTR, is held to diminish the energy consumption at the peak periods.

The contributions of this paper are summarized as follows:
\begin{itemize}
	\item A novel incentive-based DR model is proposed. Demand function is formulated by using a sigmoid function between two linear polynomials
	to depict the energy threshold when DR is required. This formulation
	is a continuous function with finite marginal value in the demand curve.
	In particular, it is an alternative modeling of DR to \citep{Su2009},
	where, the demand curve has two parts: perfect inelastic behavior
	and price responsive consumers. The inconvenience of \citep{Su2009}
	is that demand does no have perfect inelastic role since the consumers
	have a limited willingness to pay. 
	\item A Nash-Cournot equilibrium is formulated as a complementary problem
	in the presence of DR \citep{Gabriel2013}. The generators compete
	without a centralized program. Cournot gaming is compared when an electricity market operates normally and when an incentive-based DR is active during peak times.
\end{itemize}
The rest of this paper is organized as follows. Section II introduces
the agent models in an electricity market. Section III, the problem formulation as Cournot Competition in the presence of DR is developed. Numerical results are described in Section IV. Discussion is presented in Section V. Finally, conclusions
are drawn in Section VI.

\section{Agent models}
In this section, a novel demand model is proposed for studying an incentive-based DR program within an electricity market. This formulation illustrates the wholesale market behavior during a day. In addition, generator models are posed under Cournot Competition.     

\subsection{Demand response model}

The most important decision unit of microeconomic theory is the demand
\citep{Varian1992,mas1995microeconomic}. In this section, a new approach
for modeling the demand is posed when consumers participate in an incentive-based DR
program. Let $T=\left\{ 1,2,...,n_{t}\right\}$ be the set of periods to take into account in the horizon time, where $n_{t}$ is last hour, that is, $n_t=24$. An aggregated demand is considered for this DR rebate model. The decision-maker's preferences are specified by giving smooth utility function
$G(q_{t})$, where $q_{t}$ is the energy consumption at time $t$.
$G(q_{t})$ depicts the level of satisfaction obtained by the demand
as a function of the total power consumption.The utility function
satisfies the following properties as proposed in \citep{Chen2012,6266724,Fahrioglu2000,VegaRedondo2003,Osborne1995}:

Property 1: $G(q_{t})$ is assumed as a concave function with respect to $q_{t}$. This implies that the marginal benefit of users is a nonincreasing function. 

\[
\frac{d^{2}G(q_{t})}{dq_{t}^{2}}\leq0\quad\forall t\in T
\]

Property 2: The marginal benefit is nonnegative.

\[
\frac{dG\left(q_{t}\right)}{dq_{t}}\geq0\quad\forall t\in T
\]

Property 3: $G(q_{t})$ is zero when the consumption level is zero.

\[
G\left(0\right)=0\quad\forall t\in T
\]

The market price is $p_{t}^{*}$ at the time $t$. The superscript
star indicates the equilibrium price. For each generator, the cost function is assumed
increasing with respect to the total energy production capacity. In addition,
the cost function is strictly convex. Then, other definitions are
considered as follows. 
\begin{definition}
	\label{def:1}The demand energy total cost is $\pi(q_{t})=p_{t}^{*}q_{t}$. 
\end{definition}

\begin{definition}
	$G(q_{t})$ is approximated by a second order polynomial around
	$\overline{q}_{t},\quad\forall t\in T$. In general, a quadratic function
	is considered. 
	
	\[
	G\left(q_{t}\right)=-\frac{\gamma_{t}}{2}\left(q_{t}-\overline{q}_{t}\right)^{2}+p_{t}^{*}\left(q_{t}-\overline{q}_{t}\right)+k\quad\forall t\in T
	\]

	being $k=\overline{q}_{t}\left(\frac{\gamma_{t}}{2}\overline{q}_{t}+p_{t}^{*}\right)$
	a constant value, obtained by Property 3. 
\end{definition}

\begin{definition}
	\label{def:The-payoff-function}The payoff function is defined as
	$U_{t}\left(q_{t}\right)=G(q_{t})-\pi(q_{t})$, which indicates the
	user benefit of consuming $q_{t}$ energy during the interval $t$. 
\end{definition}

Basically, incentive-based DR programs request customers for curtailing demand in response
to a price signal or economic incentive. Typically the invitation
to reduce demand is made for a specific time period or peak event.
There are some concepts in order to define DR rebate program: 

\begin{definition}
	Baseline ($\beta_t$): the amount of energy the user would have consumed in the
	absence of a request to reduce (counterfactual model) \citep{Deng2015}.
	This quantity can not be measured, then this is estimated from the previous
	consumption of the agent. In this work, the aggregated baseline corresponds to the sum of all customer baseline loads in order to propose a DR threshold required in the electricity market.      
\end{definition}

$q_{t}$ is the actual use, namely, the amount of energy that aggregated demand actually consumes during the event period.

\begin{definition}
	Load Reduction ($\triangle_{t}\left(\beta_t,q_{t}\right)$): the
	difference between the baseline and the actual use. 
	
	\[
	\beta_t-q_{t}=\triangle_{t}
	\]
	
\end{definition}

In incentive-based DR programs, the rebate is only received if there
is an energy reduction. Otherwise, the user does not get any incentive
or penalty. Mathematically, 

\begin{definition}
	Let $p_{2}$ be the rebate price received by the demand due to energy reduction
	in peak periods. The DR incentive $\pi_{2}$ is given by,
	
	\[
	\pi_{2}\left(q_{t};p_{2t},\beta_t\right)=\left\{ \begin{array}{cc}
	p_{2t}\triangle_{t}=p_{2t}(\beta_t-q_{t}) & q_{t}<\beta_t\\
	0 & q_{t}\geq \beta_t
	\end{array}\right.\quad\forall t\in T
	\]

Next, the demand payoff function with DR rebate program is written as: 
	\begin{equation}
	\hat{U_{t}}\left(q_{t};p_{2t},\beta_t\right)=G(q_{t})-\pi(q_{t})+\pi_{2}(q_{t};p_{2t},\beta_t)\quad\forall t\in T\label{eq:ut}
	\end{equation}
	
\end{definition}

In this paper, the inverse demand function is formulated to develop
the Cournot\textquoteright s model of oligopoly. The inverse demand
function is given by $p_{t}\left(q_{t}\right)=\frac{dU_{t}\left(q_{t}\right)}{dq_{t}}$.
Where $p_{t}\left(q_{t}\right)$ is the price function at
the time $t$. 

Accordingly, the inverse demand function without DR is obtained from
definition \ref{def:The-payoff-function}. Next, the linear inverse demand
function is derived as follows.

\begin{equation}
p_{t}\left(q_{t}\right)=-\gamma_{t}q_{t}+\gamma_{t}\overline{q_{t}}\quad\forall t\in T\label{eq:inversedemandfunction}
\end{equation}

Whether the demand payoff function with DR is considered when $q_{t}<\beta_t$, then the inverse demand function is given by,

\begin{equation}
\widetilde{p_{t}}\left(q_{t}\right)=-\gamma_{t}q_{t}+\left(\gamma_{t}\overline{q_{t}}-p_{2t}\right)\quad\forall t\in T\label{eq:inversedemandfunction-1}
\end{equation}

\begin{figure}[h]
	\begin{centering}
		\includegraphics[scale=1]{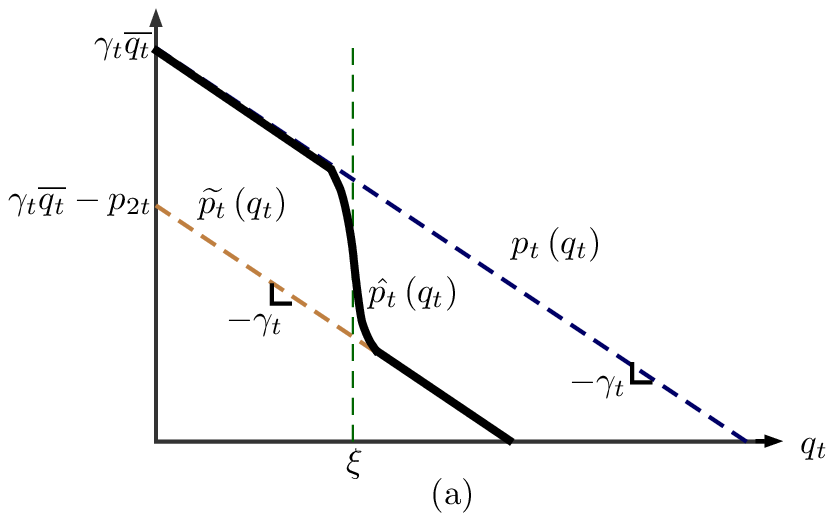}\\ \includegraphics[scale=1]{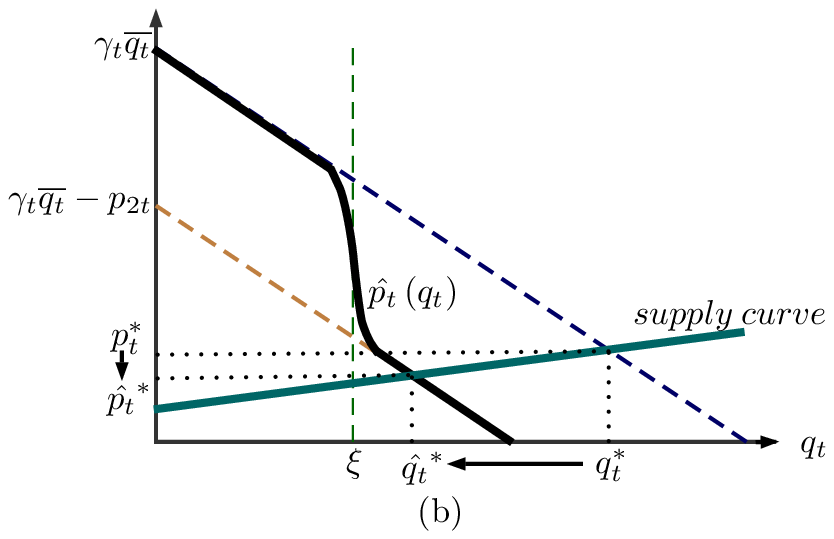}
		\par\end{centering}
	
	\caption{\label{fig:Inverse-demand-function}Inverse demand function.}
\end{figure}

In order to model the electricity market with DR during peak hours, a sigmoid function between both inverse demand functions (Eq. (\ref{eq:inversedemandfunction}) and Eq. (\ref{eq:inversedemandfunction-1})) is proposed. Fig. \ref{fig:Inverse-demand-function} (a) depicts a novel demand function that models incentive-based DR at market level. The novel inverse demand function is presented as follows. 
\begin{equation}
\hat{p_{t}}\left(q_{t}\right)=-\gamma_{t}q_{t}+\left(\gamma_{t}\overline{q_{t}}-\frac{p_{2t}}{1+e^{\alpha(-q_{t}+\xi)}}\right)\quad\forall t\in T\label{eq:inverse demand function with DR}
\end{equation}

where $\alpha$ is a constant value which represents the smoothness of the sigmoid function that joins the two straight lines and $\xi$ is the threshold level to perform the DR process.\\ 
Notice that this demand model represents a preference alteration of consumers.
Fig \ref{fig:Inverse-demand-function} (b) shows the case when the
supply curve is intersected by demand curve for $q_{t}>\beta_t$. The equilibrium
price $\hat{p_{t}}^{*}$ is less than the energy price given by the
inverse demand function $p_{t}\left(q_{t}\right)$. In addition, the
energy consumption decreases to $\hat{q_{t}}^{*}$ owing to the incentive
price $\pi_{2}$, which is requested by setting
the threshold $\xi$. The incentive is paid to consumers when the DR program
is required. Besides, SO determines the threshold $\xi$ according to
the available energy (water reservoirs, fuels, etc.), the estimated baseline $\beta_t$ and the energy consumption patterns. Then aggregators encourage customers for carrying out the energy reduction.

In \citep{Su2009}, the demand model has two parts: consumers that
have perfect inelastic behavior, they are represented by an infinite
marginal value; and users that participate in a DR program, they can place
bid price with a finite marginal value in the demand curve.
However, the drawback of the proposal \citep{Su2009} is that demand
does not have perfect inelastic behavior because the consumers have
a limit willingness for energy payment. In this paper, demand
always has a finite marginal value, hence, this model is an alternative to represent the DR behavior in an electricity market.

\subsection{Supply model}

The relationship between total energy from all generators and price
lead to make the supply curve. In this work, producers try to anticipate
the results of their actions on the price, then the market
experiences imperfect competition. SO has arbitrage services, commands DR threshold, and manages
the transmission assets as its functions into the electricity market.
Therefore, each generator seeks independently to maximize its own
economic benefits. The profit is given by its revenue from sold energy
minus the cost of generating it. Two kinds of power suppliers are
considered: thermal generators are represented with quadratic costs
and hydropower are formulated with fixed costs \citep{Genc2008}.

\subsubsection{Thermal generation modeling }

The thermal cost is given by an increasing quadratic function. Let
$r_{ta}$ be the power generated by producer $a\in A$ at the period
$t$, where $A$ is the set of thermal generators. Thus, the costs
have the following form: $c1_{a}r_{ta}+\frac{c2_{a}}{2}r_{ta}^{2}+c3_{a}$,
being $c1_{a}$, $c2_{a}$ and $c3_{a}$ constant values that depict
private information. These quadratic costs are stated because thermal
power has an expensive economic behavior \citep{Genc2008}. In this
sense, each generator uses its knowledge of the inverse demand function
($p_{t}\left(q_{t}\right)$ or $\hat{p_{t}}\left(q_{t}\right)$) to
anticipate its own effect on the market price in order to maximize
its profit. Then, the optimization problem for thermal generator is
posed as follows.

\begin{equation}
\begin{array}{c}
\max\;\sum_{t\in T}\left[p_{t}\left(q_{t}\right)r_{ta}-\left(c1_{a}r_{ta}+\frac{c2_{a}}{2}r_{ta}^{2}+c3_{a}\right)\right]\\
\begin{array}{cc}
\mathrm{s.t.}\: & r_{ta}\leq r_{a}^{+}\,:\,\mu_{ta}^{T}\quad\forall t\in T\\
& r_{ta}\geq0\quad\forall t\in T
\end{array}
\end{array}\label{eq:thermal}
\end{equation}

where $r_{a}^{+}$ is the maximum value of the energy that each thermal
power can generate in each period. $\mu_{ta}^{T}$ is dual variable
for the first constraint. This model does not consider ramp constraint,
minimum uptime and downtime, among other constraints.

\subsubsection{Hydropower modeling}

Hydropower is included in competition into the electricity
market. The hydro generator has a production function $H_{tb}(w_{tb})$ which represents the conversion of water release to energy, where $w_{tb}$ is the water release
of hydro reservoir for each generator $b\in B$. For this kind of
producer, a fixed cost $c4_{b}$ is formulated. Hence, the optimization
problem is to maximize the profits by each hydropower. 

\begin{equation}
\begin{array}{c}
\max\;\sum_{t\in T}\left[p_{t}\left(q_{t}\right)H_{tb}(w_{tb})-c4_{b}\right]\\
\begin{array}{cc}
\mathrm{s.t.\:} & w_{tb}\leq w_{tb}^{+}\,:\,\mu_{tb}^{H}\:\forall t\in T\\
& w_{tb}\geq0\:\forall t\in T
\end{array}
\end{array}\label{eq:hydro}
\end{equation}

where $w_{tb}^{+}$ is the maximum value of the water release at the
time $t$ for the generator $b$. $\mu_{tb}^{H}$ is dual variable
for the first constraint.

\section{Incentive-based demand response in Cournot competition}

A Cournot competition is developed for studying the proposed DR model that
is described in section 2. This model assumes that generators cannot
collude or form a cartel, and they seek to maximize their own profit
based on demand model. This section describes the game between market
participants in order to settle the energy price by solving simultaneously
the optimization problems (\ref{eq:thermal}) and (\ref{eq:hydro}),
as presented in \citep{Gabriel2013}. Now, the definition
of Nash equilibrium is stated as follows. 
\begin{definition}
	Considering the game $G=\left\langle I,\left\{ S_{i}\right\} _{i=1,2,..n_{i}.},\left\{ \psi_{i}\right\} _{i=1,2,...,n_{i}.}\right\rangle $,
	where $I$ is the players set, $S_{i}$ is the strategies set of each
	player and $\psi_{i}\,:\,\prod_{i\in I}S_{i}\rightarrow \mathbb{R}$ is the utility function
	of each generator. $\left(s_{1}^{*},...s_{i}^{*}\right)$ is a
	Nash equilibrium whether $\forall i \in I$ player is true that: $\psi_{i}\left(s_{i}^{*},s_{-i}^{*}\right)\geq\psi_{i}\left(s_{i},s_{-i}^{*}\right)$,
	$\forall s_{i}\in S_{i}$, being $s_{-i}$ all strategies except the
	player $i$ \citep{VegaRedondo2003}. \end{definition}
\begin{remark}
	Nash equilibrium has two interpretations: $s_{i}^{*}$ is the best
	response to $s_{-i}^{*}$ or it does not exist unilateral incentives
	to deviate from Nash equilibrium. Furthermore, an equilibrium problem
	can be solved using Karush-Kuhn-Tucker (KKT) conditions of several interrelated optimization
	problem \citep{Gabriel2013}.
	
	First, the aim is to solve Eq. (\ref{eq:thermal}) and Eq. (\ref{eq:hydro})
	in the case when no demand response is required, i.e, using the
	demand model $p_{t}\left(q_{t}\right)$ given by Eq. (\ref{eq:inversedemandfunction}). Next, the situation when DR is requested, Eq. (\ref{eq:inverse demand function with DR}) is used as demand model given by the threshold defined by SO. In order to find the solution, the
	KKT conditions of each agent are solved simultaneously. In particular, if DR is applied then the demand side shifts its energy requirement during the day in order to maintain its preferences and satisfaction levels, therefore, balance constraints are included in this case to model this behavior, namely, $\sum_{t\in T}r_t+H_t=D_n$ is added to the optimization problem, being $D_n$ the estimated net demand without DR. 
	
	In this paper, a duopoly is assumed for understanding the effect of the proposed
	DR model. In particular, two generators are employed to find the Nash-Cournot
	equilibrium: one thermal energy producer and one hydropower according
	to the suggested supply curve from Section 3. For simplicity, the
	subscript $a$ and $b$ from Eq. (\ref{eq:thermal}) and Eq. (\ref{eq:hydro})
	are removed because there is one generator per technology. Therefore,
	the net energy consumption is $q_{t}=r_{t}+H_{t}(w_{t})$ for each $t \in T$. The KKT conditions with and
	without DR are presented in the next Section. 
\end{remark}

\subsubsection{Electricity market without demand response}

First, considering the case when DR is not required in the market.
The KKT conditions are rewritten as complementary model by using
Eq. (\ref{eq:inversedemandfunction}) which are shown below.

\begin{equation}
\begin{array}{c}
0\leq r_{t}\left(2\gamma_{t}+c_{2}\right)+\left(\gamma_{t}H_{t}\left(w_{t}\right)+c_{1}\right)-\gamma_{t}\overline{q_{t}}+\mu_{t}^{T}\,\perp\,r_{t}\geq0\quad\forall t\in T\\
0\leq\mu_{t}^{T}\,\perp\,r^{+}-r_{t}\geq0\quad\forall t\in T
\end{array}\label{eq:KKTtermal}
\end{equation}
\begin{equation}
\begin{array}{c}
0\leq\frac{dH_{t}\left(w_{t}\right)}{dw_{t}}\left[\gamma_{t}r_{t}+2\gamma_{t}H_{t}\left(w_{t}\right)-\gamma_{t}\overline{q_{t}}\right]+\mu_{t}^{H}\,\perp\,w_{t}\geq0\quad\forall t\in T\\
0\leq\mu_{t}^{H}\,\perp\,w_{t}^{+}-w_{t}\geq0\quad\forall t\in T
\end{array}\label{eq:KKThydro}
\end{equation}

where (\ref{eq:KKTtermal}) and (\ref{eq:KKThydro}) are the resulting conditions
for thermal generation and hydropower,
respectively. Note that Eq. (\ref{eq:KKTtermal}) and Eq. (\ref{eq:KKThydro}) do not have interconnected periods since each hour of a day has energy consumption requirement which is depicted by an independent demand model.

\subsubsection{Electricity market with demand response}

Next, the KKT conditions are presented as follows when the market has an incentive command
by reducing energy consumption given by the demand model from Eq. (\ref{eq:inverse demand function with DR}).

\begin{equation}
\begin{array}{c}
0\leq l_r+p_{2t}\frac{e^{\alpha(r_{t}+H_{t}\left(w_{t}\right))}\left[e^{\alpha(r_{t}+H_{t}\left(w_{t}\right))}+e^{\alpha \xi}\left(\alpha r_{t}+1\right)\right]}{\left[e^{\alpha(r_{t}+H_{t}\left(w_{t}\right))}+e^{\alpha \xi}\right]^{2}}+r_{t}\left(2\gamma_{t}+c_{2}\right)\\
+\left(\gamma_{t}H_{t}\left(w_{t}\right)+c_{1}\right)-\gamma_{t}\overline{q_{t}}+\mu_{t}^{T}\,\perp\,r_{t}\geq0\quad\forall t\in T\\
0\leq\mu_{t}^{T}\,\perp\,r^{+}-r_{t}\geq0\quad\forall t\in T\\
\sum_{t\in T}r_t+H_t=D_n, \quad l_r \; \textrm{free} 
\end{array}\label{eq:termalKKTDR}
\end{equation}
\begin{equation}
\begin{array}{c}
0\leq l_h+\frac{dH_{t}\left(w_{t}\right)}{dw_{t}}\left[p_{2t}\frac{e^{\alpha(r_{t}+H_{t}\left(w_{t}\right))}\left[e^{\alpha(r_{t}+H_{t}\left(w_{t}\right))}+e^{\alpha \xi}\left(\alpha H_{t}\left(w_{t}\right)+1\right)\right]}{\left[e^{\alpha(r_{t}+H_{t}\left(w_{t}\right))}+e^{\alpha \xi}\right]^{2}}\right]\\
+\frac{dH_{t}\left(w_{t}\right)}{dw_{t}}\left[\gamma_{t}r_{t}+2\gamma_{t}H_{t}\left(w_{t}\right)-\gamma_{t}\overline{q_{t}}\right]+\mu_{t}^{H}\,\perp\,w_{t}\geq0\quad\forall t\in T\\
0\leq\mu_{t}^{H}\,\perp\,w_{t}^{+}-w_{t}\geq0\quad\forall t\in T\\
\sum_{t\in T}r_t+H_t=D_n, \quad l_h \; \textrm{free}
\end{array}\label{eq:hydroKKTDR}
\end{equation}

where (\ref{eq:termalKKTDR}) and (\ref{eq:hydroKKTDR}) are the KKT
conditions for thermal generation and hydropower if DR is applied,
respectively. $l_r$ and $l_h$ are the dual variables associated to balance constraints for thermal generation and hydropower, correspondingly. For this case, a balance constraint between all periods is added to model the shift in energy load that consumers perform to maintain their activities or their comfort levels during a day.

\section{Numerical results}

The analysis of numerical examples involves three aspects: the effect of demand response,
the study of consumer and generator surplus and the effect on the incentive variation.
The simulation is performed in GAMS 24.7.4 using PATH as the solver.\\
In Table \ref{tab:Parameters-for-simulation}, the simulation parameters are shown in order to illustrate the new approach of demand model
with DR given by Eq. (\ref{eq:inverse demand function with DR}). The simulation data are based on \citep{Forouzandehmehr2014,Genc2008,Cunningham2002}.

\begin{table}[h]
	\begin{centering}
		\begin{tabular}{cc}
			\toprule 
			$T=\left\{ 1,2,3,...,24\right\} \:h$ & $\xi=1000\:MWh$, $H_{t}(w_{t})=w_{t}$, $\alpha=0.1$\tabularnewline
			\midrule
			\midrule 
			$c1=10\:\$$ & $\begin{array}{c}
			\gamma_{t}=\{0.065,0.067,0.063,0.063,0.06,\\
			0.065,0.062,0.068,0.065,0.067,0.063,0.067,\\
			0.068,0.069, 0.062,0.061, 0.067,0.067,0.055,\\
			0.054,0.055,0.065,0.063,0.061\}\:\frac{\$}{MWh^2}
			\end{array}$\tabularnewline
			\midrule
			\midrule 
			$c2=0.025\:\$$ & $\begin{array}{c}
			\gamma_{t}\overline{q_{t}}=\{92.4, 93.82, 95.67, 99.2, 95.32, 94.56,\\ 
			90.56, 91.14, 90.19, 92.23, 91.45, 95.7,\\ 
			104.45, 103.13, 101.54, 91.87, 103.95, 95.23,\\
			120.19, 120.35, 120.23, 108.4, 95.67, 95.67\}\:\frac{\$}{MWh}
			\end{array}$\tabularnewline
			\midrule
			\midrule 
			$c3=0\:\$$ & $\begin{array}{c}
			p_{2t}=\{0,0,0,0,0,0,0,0,\\
			0,0,0,0,0,0,0,0,\\
			0,0,20,20,20,0,0,0\}\:\frac{\$}{MWh}
			\end{array}$\tabularnewline
			\midrule
			\midrule 
			$c4=0\:\$$ & $r^{+}=500\:MWh$, $w_{t}^{+}=1000\:\frac{acre-ft}{h}\:\forall t\in T$\tabularnewline
			\bottomrule
		\end{tabular}
		\par\end{centering}
	
	\caption{\label{tab:Parameters-for-simulation}Simulation parameters. }
\end{table}

\begin{figure}[h]
	\begin{centering}
		\includegraphics[scale=0.8]{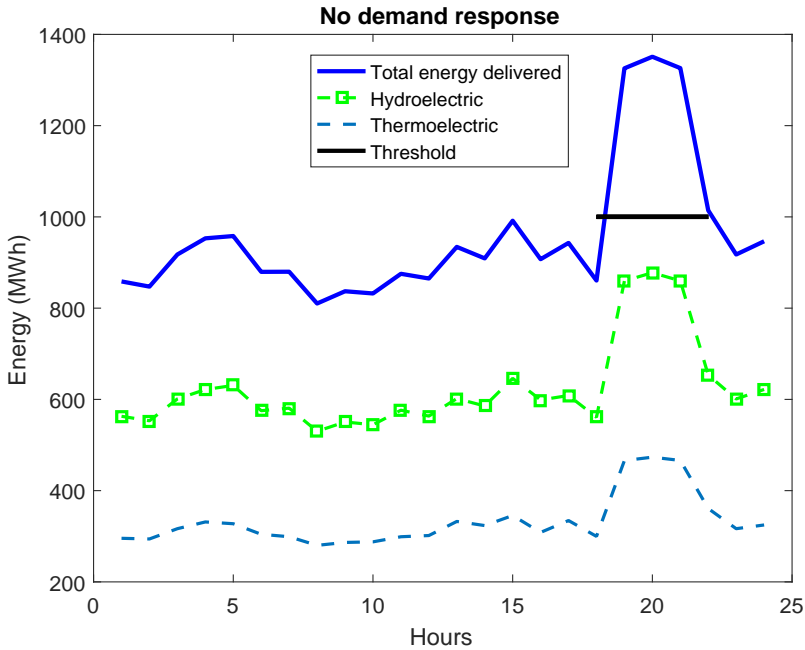}
		\par\end{centering}
	
	\caption{\label{fig:Cournot-competition-without}Cournot competition without
		demand response.}
\end{figure}

First, the Cournot competition between generators without DR is shown. Fig. \ref{fig:Cournot-competition-without} depicts
the results of gaming between thermoelectric and hydroelectric when
the inverse demand function is given by Eq. (\ref{eq:inversedemandfunction}).
The equilibrium energy versus hours in a day are depicted in
Fig. \ref{fig:Cournot-competition-without} according to the generator
technology and the total electrical energy delivered to customers. Simulations are made in a 24-hour horizon. Hydropower
has the main participation in the market due to it does not have the
variable cost, therefore, it is cheaper than the thermal generation.
The peak time occurs between 19 to 21 hours. Lastly, the net demand for all periods without DR is $D_n=25476.4\:MWh$.

\begin{figure}[h]
	\begin{centering}
		\includegraphics[scale=0.8]{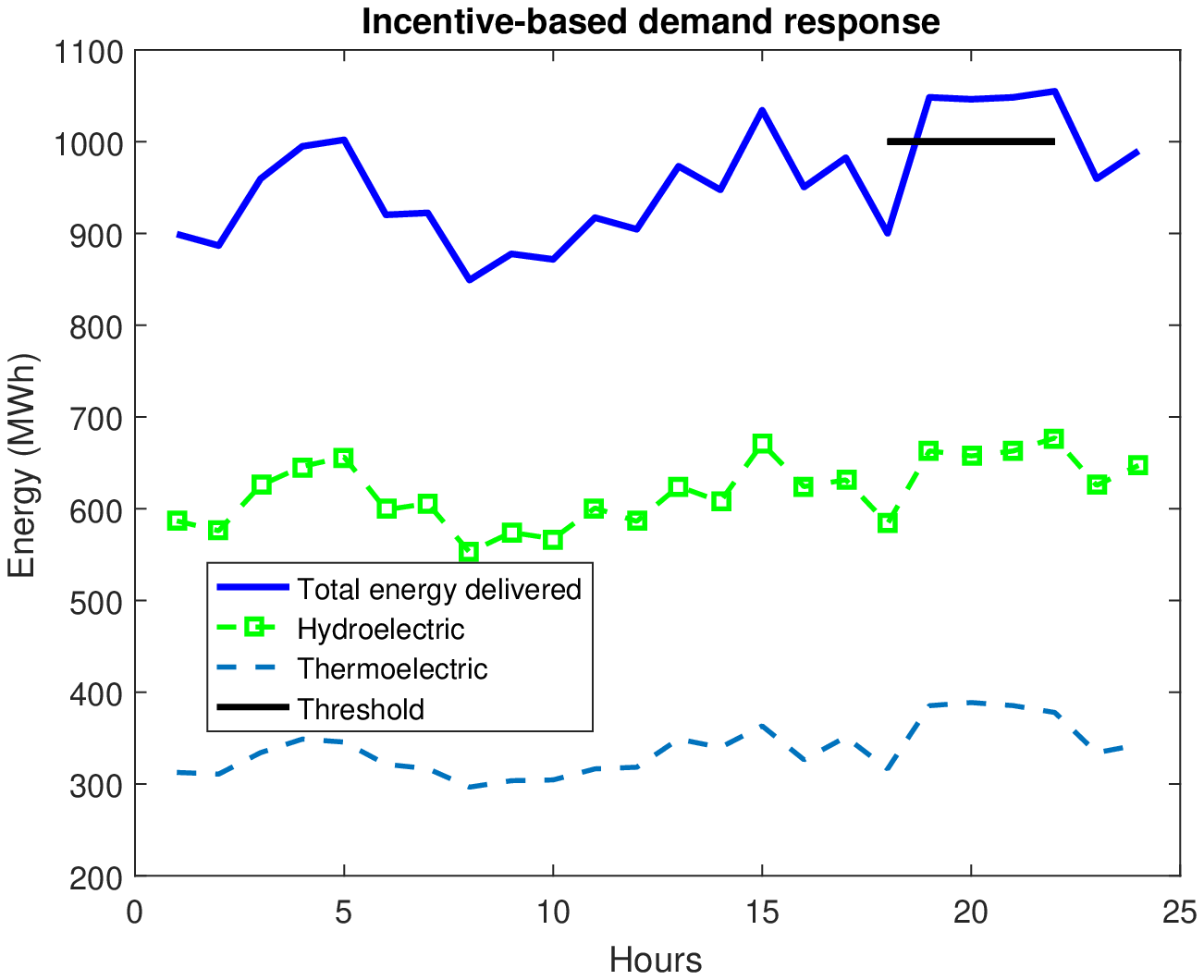}
		\par\end{centering}
	
	\caption{\label{fig:Cournot-competition-with}Cournot competition with demand
		response.}
\end{figure}

Next, Fig. \ref{fig:Cournot-competition-with} presents the competition case when there is an incentive command if the energy
consumption is greater than the threshold  $\xi=1000$ $MWh$.
Below this value, the DR benefits do not apply. For instance, notice
that the total energy delivered at the hour 20 is about 1351 $MWh$ in Fig. \ref{fig:Cournot-competition-without},
i.e., above the baseline. As long as, in Fig. \ref{fig:Cournot-competition-with},
the energy value at the same time is around 1046 $MWh$, therefore,
the energy reduction is approximately 305 $MWh$ since the demand behavior
is altered by the incentive payment given by the definition 6. In
addition, the reduction proportion is similar for each technology. For this case,  $D_n=25476.4\:MWh$ is used as estimated demand to solve Eq (\ref{eq:termalKKTDR}) and Eq. (\ref{eq:hydroKKTDR}). Hence, the net demand is the same for both situations. If DR is employed, consumers shift energy consumption to an off period in order to maintain their satisfaction level.

\subsection{The effect of demand response}

\begin{figure}[h]
	\begin{centering}
		\includegraphics[scale=0.8]{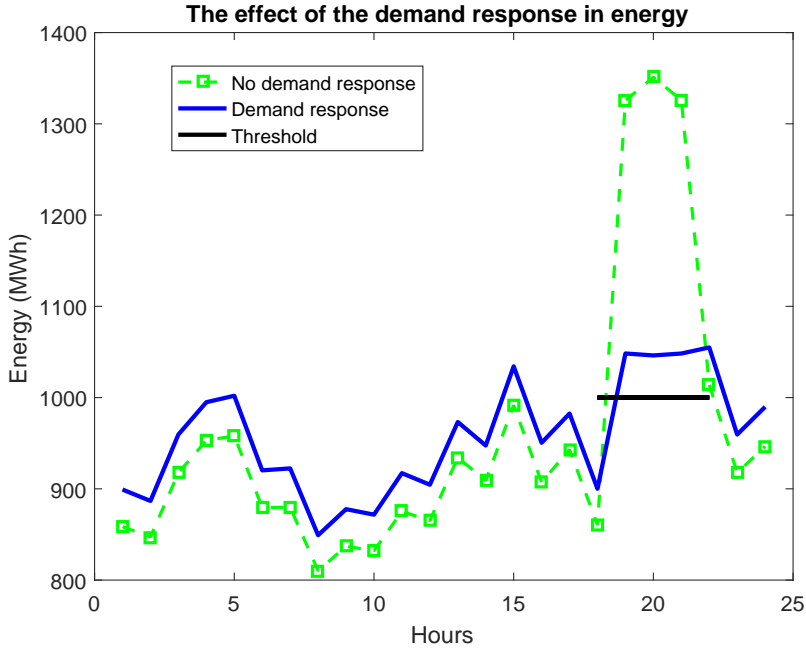}
		\par\end{centering}
	
	\caption{\label{fig:The-effect-of energy}The effect of the demand response
		in energy.}
\end{figure}

In Fig. \ref{fig:The-effect-of energy}, the effect of
DR in terms of energy is shown. Whether the consumption is higher than the threshold
value ($\xi=1000\:MWh$) and if the period has reduction incentive then the DR model stimulates the consumers
to reduce the energy consumption patterns. This behavior is found because the economic incentive $p_{2t}$ is introduced on the inverse demand function.
Thus, this incentive payment can be understood as an alteration of consumer preferences
made by SO, to alleviate the system in contingency situations
where an energy reduction is required in the grid operation. For this example,
the cutback during peak times is about 21.5\%. This percentage
changes according to the threshold selected by SO. Furthermore, demand shifts the energy requirement to other periods to hold the same activities during the day by increasing energy consumption at off-peak times.    

\begin{figure}[h]
	\begin{centering}
		\includegraphics[scale=0.8]{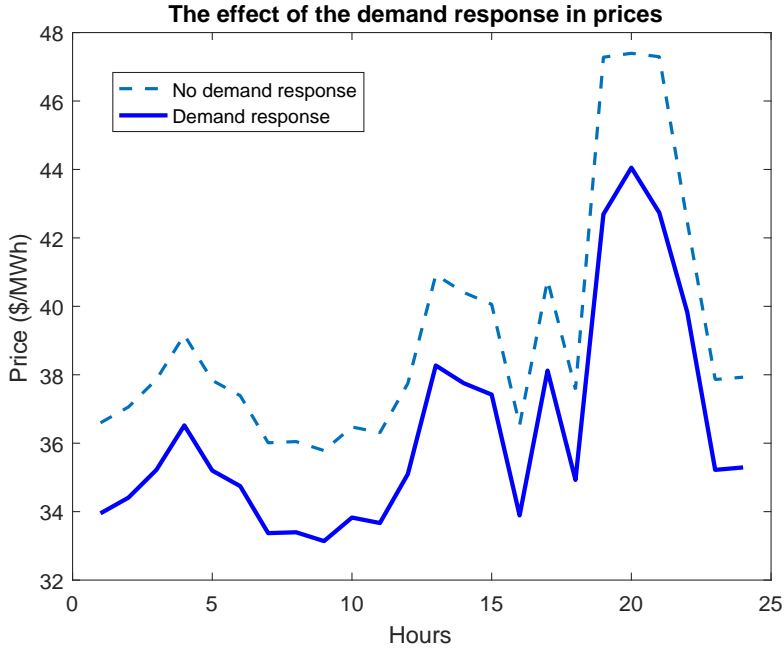}
		\par\end{centering}
	
	\caption{\label{fig:The-effect-of prices}The effect of demand response in
		prices.}
\end{figure}

In Fig. \ref{fig:The-effect-of prices}, the effects of the DR in term of prices is depicted. The fashion in which a consumer
reduces his energy is through economic stimulus or incentives.
Under this program, consumers are rewarded by a reduction of load
in peak hours. Fig. \ref{fig:The-effect-of prices}
shows that DR reduces the market prices since obeying the law
of supply and demand for all periods. However, for obtaining the energy reduction,
SO must pay an economic incentive in order to motivate the load curtailment
by consumers. Therefore, in certain events, the incentive-based DR
is a reasonable alternative to overcome contingency scenarios in the
electric power system. At these times, it is more cost-effective to
diminish demand than to increase supply or induce power outages to maintain the balance in the electrical grid. 

\subsection{Consumer and producer surplus}

\begin{figure}[h]
	\begin{centering}
		\includegraphics[scale=0.8]{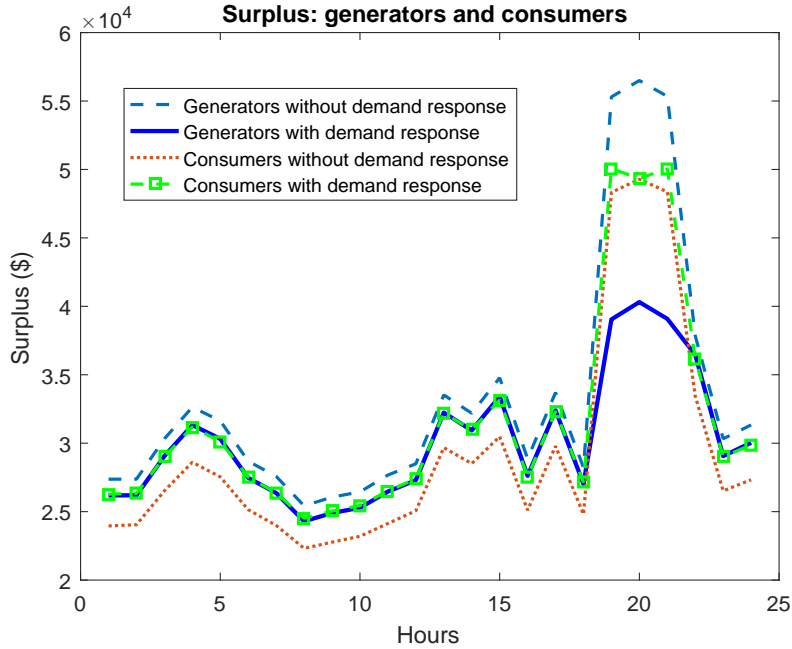}
		\par\end{centering}
	
	\caption{\label{fig:Consumer-and-producer}Consumer and producer surplus.}
\end{figure}

Consumer and generator surplus are shown in Fig. \ref{fig:Consumer-and-producer}. The producer surplus is calculated from the objective functions (Eq. (\ref{eq:thermal})
and Eq. (\ref{eq:KKThydro})). Whereas the consumer surplus is obtained
by replacing directly the inverse demand function (\ref{eq:inverse demand function with DR})
in $\int_{0}^{q_{t}}p_{t}\left(E'\right)dE'-p_{t}^{*}q_{t}$. An important feature of the
incentive-based DR program is that the generators decrease their
profit or surplus when DR is required. This effect is due to the reduction
performed by users, in which, the prices are affected by the inverse
demand curve stated when the energy exceeds the threshold. Moreover, consumers are rewarded by a reduction in their energy bill
whether they reduce their consumption. Therefore, users have a greater
economic surplus with DR program than not participating, taking into account the previous definitions for this model. 

\begin{figure}[h]
	\begin{centering}
		\includegraphics[scale=0.8]{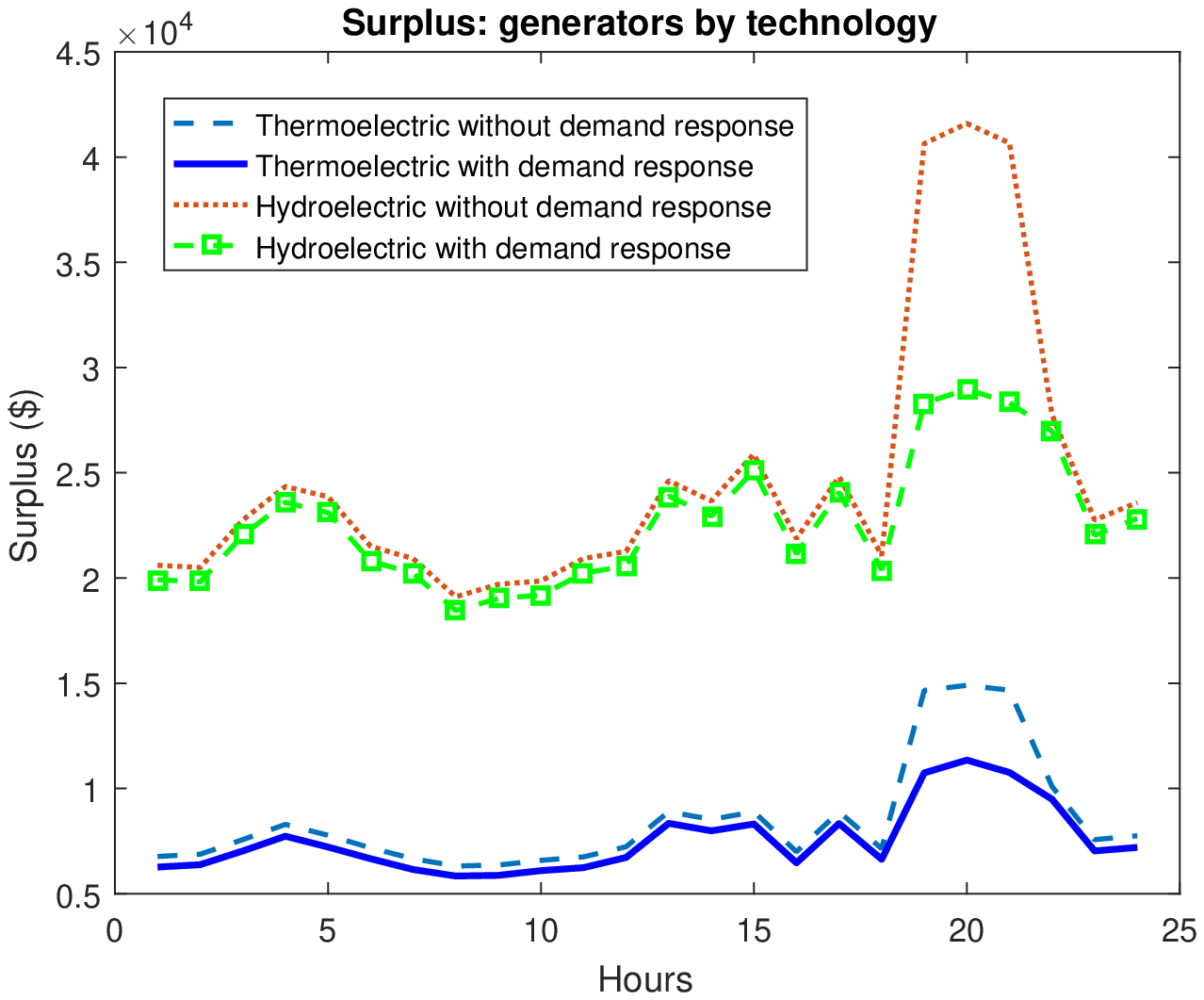}
		\par\end{centering}
	
	\caption{\label{fig:Generators-surplus-by techonology}Generators surplus by
		technology.}
\end{figure}

In Fig. \ref{fig:Generators-surplus-by techonology}, the generator surplus by technology is illustrated. Hydropower has the major participation
in the electricity market, therefore, it suffers the greatest reduction
in its benefits. Whereas that thermoelectric reduces slightly its
profit. In general, the energy reduction depends
on the participation of each generator in the energy market. For instance, 30.4\% and 23.8\% are the reduction percentages of hydroelectric and thermoelectric at 20 hour, respectively.   

\subsection{Incentive effect in demand response}

\begin{figure}[h]
	\begin{centering}
		\includegraphics[scale=0.8]{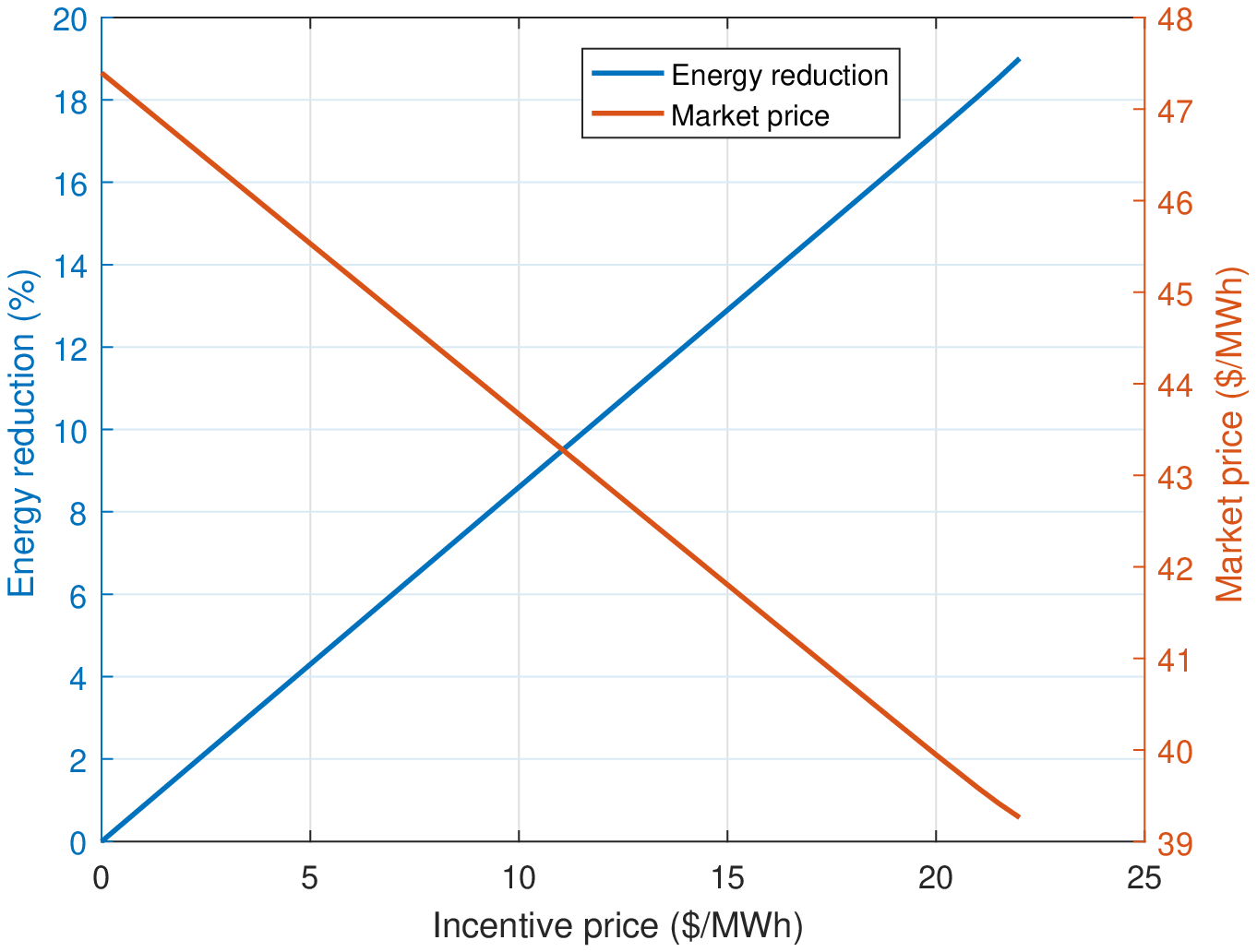}
		\par\end{centering}
	
	\caption{\label{fig:percentage-1}Energy reduction and market price affected
		by the DR incentive.}
\end{figure}

For analyzing the incentive effect in this kind of DR program, the
simulation parameters are set in $\gamma_{t}=0.054\,\$/MWh^2$
and $\gamma_{t}\overline{q_{t}}=120.35\,\$/MWh$ for one period. Next, the aim is to change the incentive price in order
to understand what happens to the energy cutback, market price,
and participant surplus. In Fig. \ref{fig:percentage-1}, the percentage energy reduction and the market price are shown
according to the incentive. The immediate effect of DR is to decrease the electric power requirement and, by the law of supply and demand, the market price declines as increases the incentive signal. For instance,
whether the incentive price $p_{2t}$ is equal to $10\:\$/MWh$, then the
market price is $43.67\:\$/MWh$ and the energy reduction is $8.6\%$. 

\begin{figure}[h]
	\begin{centering}
		\includegraphics[scale=0.8]{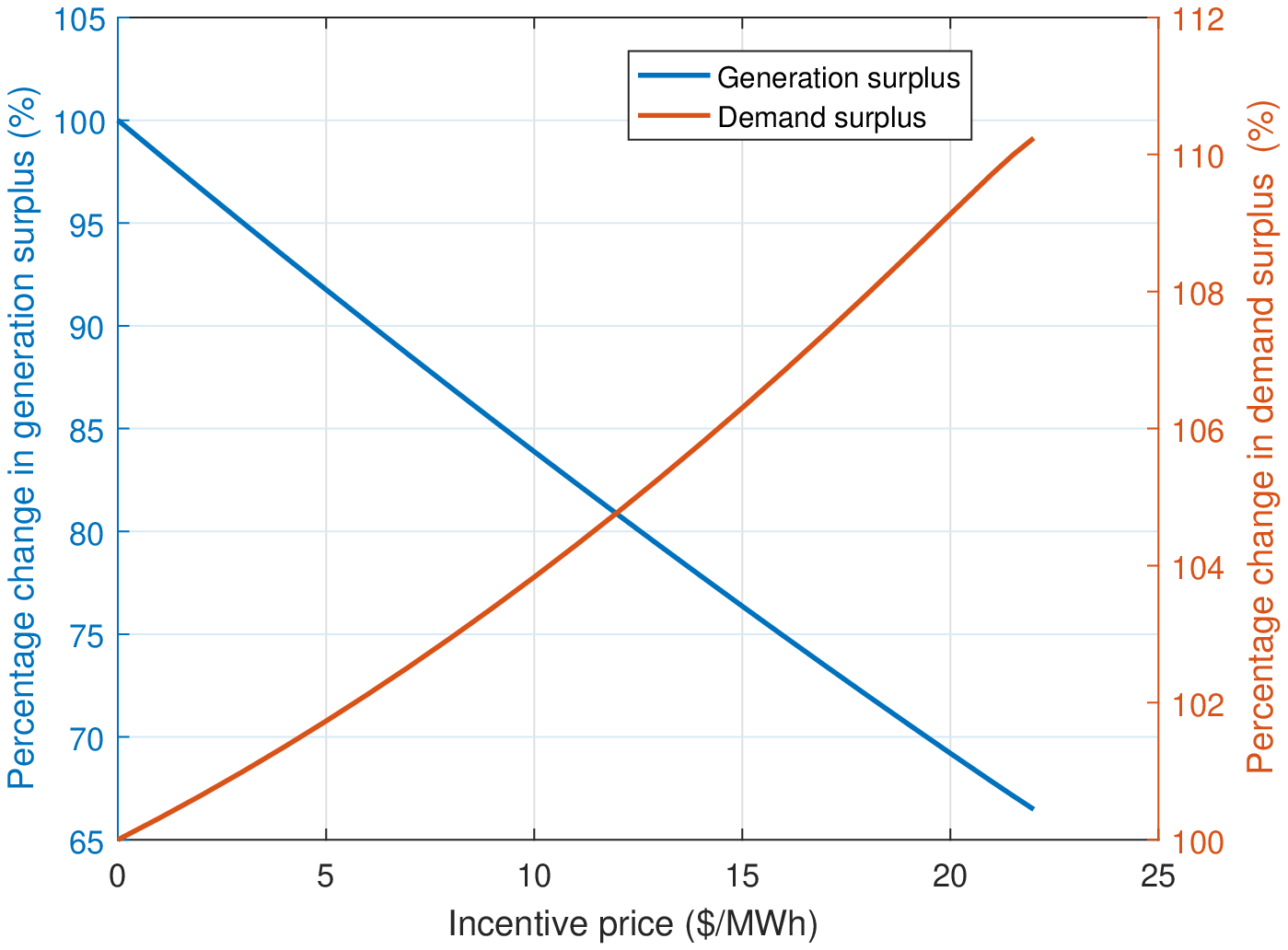}
		\par\end{centering}
	
	\caption{\label{fig:percentage-2}Generation and demand surplus behavior according
		to the incentive price}
\end{figure}

Fig. \ref{fig:percentage-2} illustrates the generation and
demand surplus as increases the incentive price. The amount of energy
to be dispatch is less when DR is required. Then, the generation profit
diminishes also caused by decreasing market prices. Therefore, the
main achievement of this incentive-based DR program is to guarantee
the power availability in peak events or to provide a solution to a
contingency situation, e.g., low water levels in reservoirs of hydroelectric
power. Furthermore, consumers perceive more economic benefits
when they are participating in the DR program since the net price
is cheaper whether they reduce their consumption. For example, if
the incentive price is 10$\$/MWh$ then users notice an increase about
3.8\% of their surplus, while, the generation has a decrease around
16.1\% of the profit. 

\section{Discussion}
Incentive-based DR is a program where consumers receive incentive payment according to a counterfactual model, namely, they are rewarded by a price multiplied by energy reduction which is measured from household baselines. Therefore, this quantity is estimated by SO from previous energy consumption. During peak times, SO evaluates the demand forecast (baseline), power availability, transmission constraints, costs of power outages, among others in order to define if DR procedure is required.\\
The proposed model allows a SO to know the market behavior in an imperfect environment to take decisions when an indirect DR method is employed. This model includes a threshold which can be interpreted as a guide or reference value of the expected energy reduction during peak time at market level. In addition, the threshold can be derived or estimated from the subset of consumers that are willing to participate in DR. This information can be collected by aggregators and analyzed by SO.\\
Economic policy is focused on how to define the incentive signal to determine a trade-off between agent surplus, grid constraints, and market objectives. For instance, under this model, the following question could be assessed: where does the incentive come from? This could be addressed by adding a fixed cost to the market price so that it can be used to encourage consumers to reduce their energy consumption in a contingency situation. Moreover, most of the literature is concentrated on directly studying DR at distribution side without considering all effects in the system as a whole. Therefore, this model provides tools to determine choices on indirect DR methods in electricity markets. \\
This new price responsive demand structure is an economic tool for analyzing DR programs in imperfect markets. In addition, this approach can be extended to study the operation of centralized systems which could result in the following benefits:  price responsive demand can make the power market more competitive during peak
times; it also can improve the predictability of
demand requirements and could provide rapid
response to emergency shortage conditions; finally, it can postpone the need for generation investment and delay the
need for certain transmission upgrades by decelerating the growth in peak demand.

\section{Conclusions}

In this work was developed an analysis of Cournot competition in an
incentive-based DR program. A new demand curve was proposed for modeling
consumer preferences in order to include DR in the electricity market.
Incentives for consumers were considered as the DR program. The demand
model was devised as a composition of two linear functions and a sigmoid,
which represents an energy threshold for analyzing the load reduction in this kind of DR programs. It was found that the incentive-based DR is a cost-effective solution
to reduce energy consumption during peak times. However, this program affects negatively the generator surplus under competition environment.\\ 
The proposed model can be employed to study price responsive demand in wholesale electricity markets where consumers have the opportunity to  reduce voluntary their consumption according to incentive signals. Particularly, price response characteristic can enable development of enhanced operational systems to take advantage of the predictable behavior of short term consumption patterns that are associated with wholesale price conditions. For instance, the model can work as decision-making tool for grid operators to defer more expensive dispatch options and reduce transmission congestion costs.\\  
For future works, transmission and intertemporal constraints can
be incorporated into the model in order to take into account all characteristic
of a dispatch problem. Additionally, a significant
improvement would be to model the demand curve as a random process for studying the electrical
grid behavior when renewable energies are integrated in the distribution system.

\begin{acknowledgements}
J. Vuelvas received a doctoral scholarship from COLCIENCIAS (Call 647-2014). This work has been partially supported by COLCIENCIAS (Grant 1203-669-4538, Acceso Universal a la Electricidad) and by Pontificia Universidad Javeriana (Grant ID 006486).
\end{acknowledgements}

\bibliographystyle{spbasic}      % basic style, author-year citations

\begin{thebibliography}{}
\bibitem[{Aketi and Sen(2014)}]{Aketi2014}
Aketi, P., Sen, S., 2014. {Modeling demand response and economic impact of
	advanced and smart metering}. Energy Systems 5~(3), 583--606.

\bibitem[{Albadi and El-Saadany(2008)}]{Albadi2008}
Albadi, M.~H., El-Saadany, E.~F., 2008. {A summary of demand response in
	electricity markets}. Electric Power Systems Research 78~(11), 1989--1996.

\bibitem[{Antunes et~al.(2013)Antunes, Faria, and Vale}]{Antunes2013}
Antunes, P., Faria, P., Vale, Z., 2013. {Consumers performance evaluation of
	the participation in demand response programs using baseline methods}. 2013
IEEE Grenoble Conference 2011, 1--6.

\bibitem[{Bloustein(2005)}]{Bloustein2005}
Bloustein, E., 2005. {Assessment of customer response to real time pricing}.
Rutgers-The State University of New Jersey, Tech. Rep, 1--23.

\bibitem[{Chen et~al.(2012)Chen, Lin, Han, Yang, Safar, and Liu}]{Chen2012}
Chen, Y., Lin, W.~S., Han, F., Yang, Y.~H., Safar, Z., Liu, K. J.~R., 2012. {A
	cheat-proof game theoretic demand response scheme for smart grids}. IEEE
International Conference on Communications, 3362--3366.

\bibitem[{Cunningham et~al.(2002)Cunningham, Baldick, and
	Baughman}]{Cunningham2002}
Cunningham, L.~B., Baldick, R., Baughman, M.~L., 2002. {An Empirical Study of
	Applied Game Theory: Transmission Constrained Cournout Behaviour}. IEEE
Transactions on power systems 17~(1), 166--172.

\bibitem[{Deng et~al.(2015)Deng, Yang, Chow, and Chen}]{Deng2015}
Deng, R., Yang, Z., Chow, M.-Y., Chen, J., 2015. {A Survey on Demand Response
	in Smart Grids: Mathematical Models and Approaches}. IEEE Transactions on
Industrial Informatics 11~(3), 1--1.

\bibitem[{Diaz et~al.(2017)Diaz, Ruiz, and Patino}]{Diaz2017}
Diaz, C., Ruiz, F., Patino, D., 2017. {Modeling and control of water booster
	pressure systems as flexible loads for demand response}. Applied Energy 204,
106--116.

\bibitem[{Fahrioglu and Alvarado(2000)}]{Fahrioglu2000}
Fahrioglu, M., Alvarado, F.~L., 2000. {Designing incentive compatible contracts
	for effective demand management}. IEEE Transactions on Power Systems 15~(4),
1255--1260.

\bibitem[{Faria et~al.(2013)Faria, Vale, and Antunes}]{Faria2013}
Faria, P., Vale, Z., Antunes, P., 2013. {Determining the adjustment baseline
	parameters to define an accurate customer baseline load}. IEEE Power and
Energy Society General Meeting 2011.

\bibitem[{Forouzandehmehr et~al.(2014)Forouzandehmehr, Han, and
	Zheng}]{Forouzandehmehr2014}
Forouzandehmehr, N., Han, Z., Zheng, R., 2014. {Stochastic Dynamic Game between
	Hydropower Plant and Thermal Power Plant in Smart Grid Networks}. IEEE
Systems Journal 10~(1), 88--96.

\bibitem[{Gabriel et~al.(2013)Gabriel, Conejo, Fuller, Hobbs, and
	Ruiz}]{Gabriel2013}
Gabriel, S.~A., Conejo, A.~J., Fuller, J.~D., Hobbs, B.~F., Ruiz, C., 2013.
{Complementarity Modeling in Energy Markets}. Vol.~1. Springer New York.

\bibitem[{Garcia et~al.(2005)Garcia, Campos-Na{\~{n}}ez, and
	Reitzes}]{Garcia2005}
Garcia, A., Campos-Na{\~{n}}ez, E., Reitzes, J., 2005. {Dynamic Pricing and
	Learning in Electricity Markets}. Operations Research 53~(2), 231--241.

\bibitem[{Genc and Thille(2008)}]{Genc2008}
Genc, T.~S., Thille, H., 2008. {Dynamic Competition in
	Electricity Markets : Hydropower and Thermal Generation}. The Economics of
Energy Markets~(November).

\bibitem[{Madaeni and Sioshansi(2013)}]{Madaeni2013}
Madaeni, S.~H., Sioshansi, R., 2013. {The impacts of stochastic programming and
	demand response on wind integration}. Energy Systems 4~(2), 109--124.

\bibitem[{Mas-Colell et~al.(1995)Mas-Colell, Whinston, and
	Green}]{mas1995microeconomic}
Mas-Colell, A., Whinston, M.~D., Green, J.~R., 1995. {Microeconomic Theory}.
Oxford student edition. Oxford University Press.

\bibitem[{Mohajeryami et~al.(2016{\natexlab{a}})Mohajeryami, Doostan, and
	Schwarz}]{Mohajeryami2016}
Mohajeryami, S., Doostan, M., Schwarz, P., 2016{\natexlab{a}}. {The impact of
	Customer Baseline Load (CBL) calculation methods on Peak Time Rebate program
	offered to residential customers}. Electric Power Systems Research 137,
59--65.

\bibitem[{Mohajeryami et~al.(2016{\natexlab{b}})Mohajeryami, Doostan, and
	Schwarz}]{Mohajeryami2016b}
Mohajeryami, S., Doostan, M., Schwarz, P., 2016{\natexlab{b}}. {The impact of
	Customer Baseline Load (CBL) calculation methods on Peak Time Rebate program
	offered to residential customers}. Electric Power Systems Research
137~(October), 59--65.

\bibitem[{Osborne(1995)}]{Osborne1995}
Osborne, M.~J., 1995. {A course in game theory}, 1st Edition. MIT press,
London.

\bibitem[{Chao(2011)}]{Chao2011}
Chao, H., 2011. {Demand response in wholesale electricity markets: the
	choice of customer baseline}. Journal of Regulatory Economics 39~(1), 68--88.

\bibitem[{Severin Borenstein}(2014)]{SeverinBorenstein2014}
{Severin Borenstein}.
\newblock {Peak-Time Rebates: Money for Nothing?}, 2014.
\newblock URL
\url{http://www.greentechmedia.com/articles/read/Peak-Time-Rebates-Money-for-Nothing}.

\bibitem[{Samadi et~al.(2012)Samadi, Mohsenian-Rad, Schober, and
	Wong}]{6266724}
Samadi, P., Mohsenian-Rad, H., Schober, R., Wong, V. W.~S., sep 2012. {Advanced
	Demand Side Management for the Future Smart Grid Using Mechanism Design}.
Smart Grid, IEEE Transactions on 3~(3), 1170--1180.

\bibitem[{Siano(2014)}]{Siano2014}
Siano, P., 2014. {Demand response and smart grids - A survey}. Renewable and
Sustainable Energy Reviews 30, 461--478.

\bibitem[{Su and Kirschen(2009)}]{Su2009}
Su, C.~L., Kirschen, D., 2009. {Quantifying the Effect of Demand Response on
	Electricity Markets}. IEEE Transactions on Power Systems 24~(3), 1199--1207.

\bibitem[{Tirole(1988)}]{Tirole1988}
Tirole, J., 1988. {The Theory of Industrial Organization}. MIT press.

\bibitem[{Vardakas et~al.(2015)Vardakas, Zorba, and Verikoukis}]{Vardakas2015}
Vardakas, J.~S., Zorba, N., Verikoukis, C.~V., 2015. {A Survey on Demand
	Response Programs in Smart Grids: Pricing Methods and Optimization
	Algorithms}. IEEE Communications Surveys {\&} Tutorials 17~(1), 152--178.

\bibitem[{Varian(1992)}]{Varian1992}
Varian, H., 1992. {Microeconomics Analysis}, 3rd Edition. Norton {\&} Company.

\bibitem[{{Vega Redondo}(2003)}]{VegaRedondo2003}
{Vega Redondo}, F., 2003. {Economics and the theory of Games}. Cambridge
University Press.

\bibitem[{Villar and Rudnick(2003)}]{Villar2003}
Villar, J., Rudnick, H., 2003. {Hydrothermal market simulator using game
	theory: Assessment of market power}. IEEE Transactions on Power Systems
18~(1), 91--98.

\bibitem[{Vuelvas and Ruiz(2015)}]{Vuelvas2015}
Vuelvas, J., Ruiz, F., 2015. {Demand response: Understanding the rational
	behavior of consumers in a Peak Time Rebate Program}. in Automatic Control
(CCAC), 2015 IEEE 2nd Colombian Conference on, 1--6.

\bibitem[{Vuelvas and Ruiz(2017)}]{Vuelvas2017}
Vuelvas, J., Ruiz, F., 2017. {Rational consumer decisions in a peak time rebate
	program}. Electric Power Systems Research 143, 533--543.

\bibitem[{Wijaya et~al.(2014)Wijaya, Vasirani, and Aberer}]{Wijaya2014}
Wijaya, T.~K., Vasirani, M., Aberer, K., 2014. {When Bias Matters: An Economic
	Assessment of Demand Response Baselines for Residential Customers}. IEEE
Transactions on Smart Grid 5~(4), 1755--1763.

\bibitem[{Zhu et~al.(2013)Zhu, Sauer, and Basar}]{Zhu2013}
Zhu, Q., Sauer, P., Basar, T., 2013. {Value of demand response in the smart
	grid}. In: 2013 IEEE Power and Energy Conference at Illinois, PECI 2013. pp.
76--82.

\end{thebibliography}


\end{document}